\title{Cancer phase I trial design using drug combinations when a fraction of dose limiting toxicities is attributable to one or more agents}
\author{Jos\'e L. Jim\'enez$^{1*}$, Mourad Tighiouart$^2$, Mauro Gasparini$^1$\\
$^1$Politecnico di Torino, Turin, Italy\\
$^2$Samuel Oschin Comprehensive Cancer Institute, Los Angeles, CA, U.S.A.\\
$^*$ E-mail: jose.jimenez@polito.it}

\documentclass[12pt]{article}

\usepackage{amsmath,amssymb}
\usepackage{bbm}
\usepackage[a4paper, total={7in, 10in}]{geometry}
\usepackage{graphicx}
\usepackage{makecell}
\usepackage{bigints}
\usepackage{multirow}
 \usepackage{relsize}
\usepackage{algorithm, algpseudocode}
\usepackage{bbm}
\usepackage{multirow}
\usepackage{amsmath,amsfonts,amssymb,amsthm,epsfig,epstopdf,titling,url,array}
\usepackage{flexisym}
\usepackage{algorithm}
\usepackage{algpseudocode}
\usepackage{mathtools}

\usepackage{color}

\usepackage{ctable}

\theoremstyle{plain}

\theoremstyle{definition}

\theoremstyle{remark}

\date{}

\begin{document}
\maketitle

\begin{abstract}
Drug combination trials are increasingly common nowadays in clinical research. However, very few methods have been developed to consider toxicity attributions in the dose escalation process. We are motivated by a trial in which the clinician is able to identify certain toxicities that can be attributed to one of the agents. We present a Bayesian adaptive design in which toxicity attributions are modeled via Copula regression and the maximum tolerated dose (MTD) curve is estimated as a function of model parameters. The dose escalation algorithm uses cohorts of two patients, following the continual reassessment method (CRM) scheme, where at each stage of the trial, we search for the dose of one agent given the current dose of the other agent. The performance of the design is studied by evaluating its operating characteristics when the underlying model is either correctly specified or misspecified. We show that this method can be extended to accommodate discrete dose combinations.
\\
\\
Keywords: Attributable toxicity; Cancer phase I trials; Continual reassessment method; Copula type models; Drug combination.
\end{abstract}

\section{Introduction}
Cancer phase I clinical trials constitute the first step in investigating a potentially promising combination of cytotoxic and biological agents. Due to safety and ethical concerns, patients are sequentially enrolled in the trial, and the dose combinations assigned to subsequent patients depend on dose combinations already given to previous patients and their dose limiting toxicity (DLT) status at the end of the first cycle of therapy. 
The main objective of these trials is to estimate a maximum tolerated dose (MTD) that will be used in future efficacy evaluation in phase II/III trials. The MTD is usually defined as any dose combination $(x,y)$ that will produce DLT in a prespecified proportion $\theta$ of patients,

\begin{equation}
\mbox{Prob}(\mbox{DLT} | \mbox{ dose} = (x,y)) = \theta.
\end{equation}

The definition of DLT depends on the type of cancer and drugs under study, but it is usually defined as a grade 3 or 4 non-hematologic toxicity (see the National Cancer Institute CTCAE v4.03 for the definition of the different grades of toxicity). The pre-specified proportion of DLTs $\theta$, sometimes referred as target probability of DLT, also depends on the nature of the toxicity, but it usually take values between 0.2 and 0.4.

In the drug combination dose finding literature, designs that recommend a unique MTD (see e.g. \cite{yin2009latent, yin2009bayesian, wages2011continual, wages2011dose, shi2013escalation, riviere2014bayesian, wheeler2017modelling, mu2017new}) or multiple MTDs (see e.g. \cite{thall2003dose,wang2005two,yuan2008sequential,
braun2010hierarchical,mander2015product,tighiouart2014dose,tighiouart2016threedrugs,
tighiouart2017bayesian}) have been studied extensively. Most of these methods use a parametric model for the dose-toxicity relationship

\begin{equation}
\mbox{Prob}(\mbox{DLT} | (x, y)) = F((x,y), {\boldsymbol \xi}),
\end{equation} where $(x,y)$ represents the drug combination of two agents, $F(.)$ is a known link function, e.g. a power model or a logistic model,  and ${\boldsymbol \xi} \in R^{d}$ is a vector of $d$ unknown parameters. Non-parametric designs have been proposed in the past, both in single agent and drug combination settings \cite{mander2015product, gasparini2000curve, whitehead2010bayesian}. These designs unique assumption is monotonicity, which is imposed either through the prior distribution (see \cite{gasparini2000curve, whitehead2010bayesian}), or by choosing only monotonic contours when escalating (see \cite{mander2015product}).

Let $S$ be the set of all dose combinations available in the trial, and $C({\boldsymbol \xi})$ be the set of dose combinations $(x,y)$ such the probability of DLT equals a target risk of toxicity $\theta$. Hence,

\begin{equation} \label{mtdset1}
C({\boldsymbol \xi}) = \{ (x, y) \in S : F((x,y),{\boldsymbol \xi}) = \theta \}.
\end{equation}

Equation \eqref{mtdset1} is the traditional definition of MTD set. When $S$ is discrete, following \cite{tighiouart2017bayesian}, we can define the MTD as the set of dose combinations $(x,y)$ that satisfy 

\begin{equation}
| F((x,y),{\boldsymbol \xi}) - \theta | \leq \delta,
\end{equation} since $C({\boldsymbol \xi})$ may be empty, i.e., when the MTD is not in $S$. The threshold parameter $\delta, 0 < \delta < 1$, is pre-specified after close collaboration with the clinician.

This work is motivated by a cancer phase I trial a clinician at Cedars-Sinai Medical Center is planning. The trial involves the combination of Taxotere, a known cytotoxic agent, and Metformin, a diabetes drug, in advanced or metastatic breast cancer patients. According to the clinician, some DLTs can be attributable to either agent or both. For example, a grade 3 or 4 neutropenia can only be attributable to Taxotere and not Metformin. Furthermore, for ethical reasons, if a patient has a DLT attributable to Taxotere when treated with dose level $x_{T}$ of taxotere, then $x_{T}$ cannot be increased for the next patient in the trial (see the dose escalation restriction in Section 2.2). Very few methods have been developed to incorporate toxicity attribution in the dose escalation process. \cite{yin2009latent} proposed a design that models the joint probability of toxicity with a copula model known as the Gumbel model \cite{murtaugh1990bivariate}. This model allows the investigator to compute the probability of DLT when the DLT is exclusively attributed to one drug, the other one, or both. 
However, they require all toxicities to be attributable, which is rare in practice. \cite{wheeler2017modelling} proposed a semi-attributable toxicity design based on a trial with non-concurrent drug administration. In their design, one drug is administered at the beginning of the treatment cycle and the other drug is administered at a much later time point if and only if the patient did not experience DLT. If a DLT occurs before the second drug is administered, then the DLT is attributed to the first drug. However, if the DLT occurs after the second drug has been administered, then the DLT could be caused by any of the drugs and therefore is not attributable. \cite{iasonos2016phase} propose a method that reduces the effect of the bias caused by toxicity attribution errors by using personalized scores instead of the traditional binary DLT outcome.
\cite{lee2012two} considered the toxicity attribution problem for ruled-based designs with non-overlapping toxicities.

In this article, we propose a Bayesian adaptive design for drug combinations that allows the investigator to attribute a DLT to one or both agents in an unknown fraction of patients, even when the drugs are given concurrently. 

We define toxicity attribution as a DLT caused by one drug and not the other when the type of DLT is non-overlapping, e.g., a grade 4 neutropenia is caused by taxotere but can never occur with metformin, or when the clinician judges that a type of DLT is caused by one drug and not the other, e.g., a grade 4 diarrhea is caused by taxotere but not metformin due to the low dose level of taxotere that was given in combination even though both drugs have this side effect in common.

The relationship between the dose combinations and the risk of toxicity is modeled using the same copula model used by \cite{yin2009latent}. The design proceeds using a variation of the algorithm proposed in \cite{tighiouart2017bayesian} where cohorts of two patients are allocated to dose combinations where, at each stage of the trial, we search for the dose of one agent given the current dose of the other agent. Our approach differs from the methodologies of \cite{yin2009latent} and \cite{tighiouart2017bayesian} in three aspects; (i) a non-negative fraction of DLTs are attributable to either one or both agents, (ii) the dose combination allocated to patients uses the CRM scheme as opposed to escalation with overdose control (EWOC) approach proposed by \cite{babb1998cancer}, and (iii) if a current patient experiences DLT attributed drug $D_{1}$ at dose level $x_{D_{1}}$, then the dose level of agent $D_1$ cannot be more than $x_{D_{1}}$ for the next cohort of two patients. At the end of the trial, an estimate of the MTD curve is proposed as a function of Bayes estimates of the model parameters. Last, we show that our method can be easily adapted from a setting with continuous dose combinations to discrete dose combinations by rounding up the estimated MTD curve to the nearest discrete dose combinations.

The rest of the manuscript is organized as follows. In Section 2, we describe the model for the dose-toxicity relationship and the adaptive design to conduct the trial for continuous dose combinations. In Section 3, we study the performance of the method in terms of safety and efficiency of the estimate of the MTD set. In Section 4, we adapt our proposal to the setting of discrete dose combinations. In section 5, we conduct a model misspecification evaluation. Discussion and practical considerations of the method are discussed in Section 6.


\section{Method}
\subsection{Dose-Toxicity Model}

Let $X_{\min}, X_{\max}, Y_{\min}, Y_{\max}$, be the minimum and maximum doses available in a trial that combines drugs with continuous dose combination levels. The doses are standardized to be in a desired interval, e.g., [0.05, 0.3], so that $X_{\min} = Y_{\min} =0.05$ and $ X_{\max} = Y_{\max} = 0.3$. Let $F_{\alpha}(\cdot)$ and $F_{\beta}(\cdot)$ be parametric models for the probability of DLT of drugs $D_{1}$ and $D_{2}$, respectively. We specify the joint dose-toxicity relationship using the Gumbel copula model (see \cite{murtaugh1990bivariate}) as

\begin{equation}
\begin{split}
&\pi^{(\delta_1,\delta_2)} = \mbox{Prob}(\delta_1, \delta_2 | x, y) = F_{\alpha}^{\delta_1}(x) \left [1 - F_{\alpha}(x) \right ]^{1-\delta_1} \times \\
& F_{\beta}^{\delta_2}(y) \left [1-F_{\beta}(y)\right ]^{1-\delta_2} + (-1)^{(\delta_1+\delta_2)} F_{\alpha}(x)\left [1-F_{\alpha}(x)\right] F_{\beta}(y) \left [1 - F_{\beta}(y)\right ]  \frac{e^{-\gamma}-1}{e^{-\gamma}+1},
\end{split}
\label{dosetoxicitymodel}
\end{equation} where $x$ is the standardized dose level of drug $D_{1}$, $y$ is the standardized dose level of agent $D_{2}$, $\delta_{1}$ is the binary indicator of DLT attributed to drug $D_{1}$, $\delta_{2}$ is the binary indicator of DLT attributed to drug $D_{2}$ and $\gamma$ is the interaction coefficient. We assume that the joint probability of DLT, when one of the drugs is held constant, is monotonically increasing; that is $ \mbox{Prob}(\mbox{DLT} | x',y) \geq \mbox{Prob}(\mbox{DLT} | x,y)$ or $\mbox{Prob}(\mbox{DLT} | x,y') \geq \mbox{Prob}(\mbox{DLT} | x,y)$, where $x' > x$ and $y'>y$. A sufficient condition for this property to hold is to assume that $F_{\alpha}(\cdot)$ and $F_{\beta}(\cdot)$ are increasing functions with $\alpha > 0$ and $\beta>0$. In this article we use $F_\alpha(x) = x^\alpha$ and $F_\beta(y) = y^\beta$. Using (\ref{dosetoxicitymodel}), if the DLT is attributed exclusively to drug $D_1$, then

\begin{equation}
\label{probdltdruga}
\begin{split}
&\pi^{(\delta_1 = 1, \delta_2 = 0)} = \mbox{Prob}(\delta_1=1,\delta_2=0 | x, y) =  x^\alpha (1-y^\beta) - x^\alpha\left (1-x^\alpha\right) y^\beta \left (1 - y^\beta\right )  \frac{e^{-\gamma}-1}{e^{-\gamma}+1}.
\end{split}
\end{equation}

If the DLT is attributed exclusively to drug $D_2$, then

\begin{equation}
\label{probdltdrugb}
\begin{split}
&\pi^{(\delta_1 = 0, \delta_2 = 1)} = \mbox{Prob}(\delta_1=0,\delta_2=1 | x, y) =  y^\beta (1- x^\alpha) - x^\alpha\left (1-x^\alpha\right) y^\beta \left (1 - y^\beta\right )  \frac{e^{-\gamma}-1}{e^{-\gamma}+1}.
\end{split}
\end{equation}

If the DLT is attributed to both drugs $D_1$ and $D_2$, then

\begin{equation}
\label{probdltdrugab}
\begin{split}
&\pi^{(\delta_1 = 1,\delta_2 = 1)} = \mbox{Prob}(\delta_1=1,\delta_2=1 | x, y) =  x^\alpha y^\beta +  x^\alpha \left (1-x^\alpha\right) y^\beta \left (1 - y^\beta\right )  \frac{e^{-\gamma}-1}{e^{-\gamma}+1}.
\end{split}
\end{equation}

Equation \eqref{probdltdruga} represents the probability that $D_1$ causes a DLT and drug $D_2$ does not cause a DLT. This can happen, for example, when a type of DLT of taxotere ($D_1$), such as grade 4 neutropenia, is observed. However, this type of DLT can never be observed with metformin ($D_2$). This can also happen when the clinician attributes a grade 4 diarrhea to taxotere ($D_1$) but not to metformin ($D_2$) in the case of a low dose level of this later even though both drugs have this common type of side effect. The fact that dose level $y$ is present in equation  \eqref{probdltdruga} is a result of the joint modeling of the two marginals and accounts for the probability that drug $D_2$ does not cause a DLT. This later case is, of course, based on the clinician’s judgment. Equations \eqref{probdltdrugb} and \eqref{probdltdrugab} can be interpreted similarly.

Following \cite{yin2009latent}, it is easy to see that the total probability of having a DLT is calculated as the sum of (\ref{probdltdruga}), (\ref{probdltdrugb}) and (\ref{probdltdrugab}). Hence,

\begin{equation}
\label{probdlt}
\begin{split}
&\pi = \mbox{Prob}(\mbox{DLT} | x,y) = \pi^{(\delta_1=1,\delta_2=0)} + \pi^{(\delta_1=0,\delta_2=1)} + \pi^{(\delta_1=1,\delta_2=1)} =\\
& x^\alpha + y^\beta - x^\alpha y^\beta - x^\alpha \left (1-x^\alpha \right) y^\beta \left (1 - y^\beta\right )  \frac{e^{-\gamma}-1}{e^{-\gamma}+1}.
\end{split}
\end{equation}

We define the MTD as any dose combination $(x_*,y_*)$ such that $\mbox{Prob}(\mbox{DLT} | x_*,y_*) = \theta$. We set \eqref{probdlt} equal to $\theta$ and re-write it as a 2nd degree polynomial in $y^\beta$, and solve for the solutions. This allows us to define the MTD set $C(\alpha,\beta,\gamma)$ as

\begin{equation}
\label{mtdset}
C(\alpha,\beta,\gamma) = \left \{(x_*,y_*): y_* = \left [ \frac{-(1 - x_*^\alpha - \kappa) \pm \sqrt{(1 - x_*^\alpha - \kappa)^2 -4\kappa(x_*^\alpha - \theta)}}{2\kappa} \right ]^{\frac{1}{\beta}} \right \},
\end{equation}where\begin{equation*}
\kappa = x_*^\alpha(1 - x_*^\alpha) \frac{e^{-\gamma}-1}{e^{-\gamma} + 1}.
\end{equation*}

Let $T$ be the indicator of DLT, $T=1$ if a patient treated at dose combination $(x,y)$ experiences DLT within one cycle of therapy that is due to either drug or both, and $T=0$ otherwise. Among patients treated with dose combination $(x,y)$ who exhibit DLT, suppose that an unknown fraction $\eta$ of these patients have a DLT with known attribution, i.e. the clinician knows if the DLT is caused by drug $D_{1}$ only, or drug $D_{2}$ only, or both drugs $D_{1}$ and $D_{2}$. Let $A$ be the indicator of DLT attribution when $T = 1$. It follows that for each patient treated with dose combination $(x,y)$, there are five possible toxicity outcomes: $\{T=0\}, \{T=1,A=0\}, \{T=1, A=1, \delta_{1}=1,\delta_{2}=0\},\{T=1, A=1, \delta_{1}=0,\delta_{2}=1\} \mbox{ and } \{T=1, A=1, \delta_{1}=1,\delta_{2}=1\}$. This is illustrated in the chance tree diagram in Figure 1. Using equations \eqref{probdltdruga},\eqref{probdltdrugb},\eqref{probdltdrugab},\eqref{probdlt} and Figure 1, the contributions to the likelihood from each of the five observable outcomes are listed in Table \ref{likelihoodtable}. The likelihood function is defined as

\begin{equation}
\label{likelihoodfunction}
L(\alpha, \beta, \gamma, \eta ~ | ~ \mbox{data}) = \prod_{i=1}^n \big[ \big( \eta  \pi_{i}^{(\delta_{1_i}, \delta_{2_i})} \big)^{A_i} \big( \pi_i \ (1-\eta) \big)^{1-A_i} \big]^{T_i} (1-\pi_i)^{1-T_i},
\end{equation} and the joint posterior probability distribution of the model parameters as 

\begin{equation}
\label{posteriorparameters}
\begin{split}
&\mbox{Prob}(\alpha, \beta, \gamma, \eta ~ | ~ \mbox{data}) \propto
\mbox{Prob}(\alpha, \beta, \gamma) \times L(\alpha, \beta, \gamma ~ | ~ \mbox{data}).
\end{split}
\end{equation}

With equation (\ref{posteriorparameters}) we can easily sample and obtain MCMC estimates of $\alpha$, $\beta$, $\gamma$ and $\eta$.

\begin{figure}[H]
\caption{A chance tree illustrating the five possible outcomes we can find in a trial.}
\centering
\includegraphics[scale=0.5]{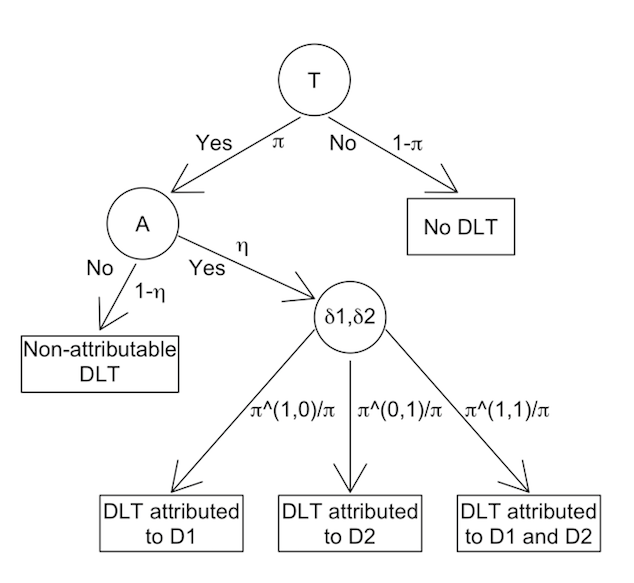}
\label{figure1}
\end{figure}

\begin{table}[H]
\caption{Contributions to the likelihood function based on the observed outcomes: toxicity, attribution, attribution to drug 1 ($\delta_1$) and attribution to drug 2 ($\delta_2$) for each patient.}
\centering
\resizebox{\columnwidth}{!}{%
\begin{tabular}{ccccc}
\toprule
\makecell{Toxicity} & \makecell{Attribution} & $\delta_1$ & $\delta_2$ & Likelihood            \\ \midrule
0   & -          & -          & -          & \makecell{ $1 - \pi = 1 - \big [ x^\alpha + y^\beta - x^\alpha \times y^\beta - $  $x^\alpha\left (1-x^\alpha\right) y^\beta \left (1 - y^\beta\right )  \frac{e^{-\gamma}-1}{e^{-\gamma}+1} \big ] $ }              \\ \midrule
1   & 0          & -          & -          & \makecell{ $ \pi \times (1-\eta) = \big [ x^\alpha + y^\beta - x^\alpha \times y^\beta - $  $x^\alpha\left (1-x^\alpha\right) y^\beta \left (1 - y^\beta\right )  \frac{e^{-\gamma}-1}{e^{-\gamma}+1} \big ] \times (1 - \eta) $ }       \\ \midrule
1   & 1          & 1          & 0          & \makecell{ $ \pi \times \eta \times \frac{\pi^{(1,0)}}{\pi} = \eta \times \big [ x^\alpha (1-y^\beta) - $  $ x^\alpha\left (1-x^\alpha\right) y^\beta \left (1 - y^\beta\right )  \frac{e^{-\gamma}-1}{e^{-\gamma}+1} \big ]$ } \\ \midrule
1   & 1          & 0          & 1          & \makecell{ $ \pi \times \eta \times \frac{\pi^{(0,1)}}{\pi} = \eta \times \big [ y^\beta (1-x^\alpha) - $  $ x^\alpha\left (1-x^\alpha\right) y^\beta \left (1 - y^\beta\right )  \frac{e^{-\gamma}-1}{e^{-\gamma}+1} \big ]$ } \\ \midrule
1   & 1          & 1          & 1          & \makecell{ $ \pi \times \eta \times \frac{\pi^{(1,1)}}{\pi} = \eta \times \big [ x^\alpha \times y^\beta + $  $ x^\alpha\left (1-x^\alpha\right) y^\beta \left (1 - y^\beta\right )  \frac{e^{-\gamma}-1}{e^{-\gamma}+1} \big ]$ } \\ \bottomrule
\end{tabular}
}
\label{likelihoodtable}
\end{table}

\subsection{Trial Design}
\label{dosefindingalgorithmsection}

Dose escalation / de-escalation proceeds using the algorithm described in \cite{tighiouart2017bayesian} but univariate continual reassessment method (CRM) is carried out to estimate the next dose instead of EWOC. In a cohort with two patients, the first one would receive a new dose of agent $D_1$ given the dose $y$ of agent $D_2$ that was previously assigned. The new dose of agent $D_1$ is defined as $x_{\small{\mbox{new}}} = \text{argmin}_u |\widehat{\mbox{Prob}}(\mbox{DLT} | u,y) - \theta |,
$where $y$ is fixed and $\hat{\mbox{Prob}}(\mbox{DLT} | u,y)$ is computed using equation (\ref{probdlt}) with $\alpha, \beta, \gamma$ replaced by their posterior medians. The other patient would receive a new dose of agent $D_2$ given the dose of agent $D_1$ that was previously assigned. Specifically, the design proceeds as follows:

\begin{enumerate}

\item Patients in the first cohort receive the same dose combination $(x_1, y_1) = (x_2, y_2)=(X_{\min}, Y_{\min})$.

\item In the $i$-th cohort of two patients,

\begin{itemize}

\item If $i$ is even, 

\begin{itemize}

\item Patient $(2i - 1)$ receives doses $(x_{2i-1},y_{2i-1})$, where
$x_{2i-1} =\underset{u}{\text{argmin}} \left |\widehat{\mbox{Prob}}(\mbox{DLT} | u,y_{2i-3}) - \theta \right |,$ and $y_{2i-1} = y_{2i-3}$. If a DLT was observed in the previous cohort of two patients and was attributable to drug $D_1$, then $x_{2i-1}$ is further restricted to be no more than $x_{2i-3}$.

\item Patient $2i$ receives doses $(x_{2i}, y_{2i})$, where $y_{2i} =\underset{v}{\text{argmin}} \left |\widehat{\mbox{Prob}}(\mbox{DLT} | x_{2i-2},v) - \theta \right |,
$ and $x_{2i} = x_{2i-2}$. If a DLT was observed in the previous cohort of two patients and was attributable to drug $D_2$, then $y_{2i}$ is further restricted to be no more than $y_{2i-2}$.

\end{itemize}

\item If $i$ is odd, 

\begin{itemize}

\item Patient $(2i - 1)$ receives doses $(x_{2i-1},y_{2i-1})$, where $y_{2i-1} =\underset{v}{\text{argmin}} \left |\widehat{\mbox{Prob}}(\mbox{DLT} | x_{2i-3},v) - \theta \right |,
$ and $x_{2i-1} = x_{2i-3}$. If a DLT was observed in the previous cohort of two patients and was attributable to drug $D_2$, then $y_{2i-1}$ is further restricted to be no more than $y_{2i-3}$.

\item Patient $2i$ receives doses $(x_{2i},y_{2i})$, where $x_{2i} =\underset{u}{\text{argmin}} \left |\widehat{\mbox{Prob}}(\mbox{DLT} | u,y_{2i-2}) - \theta \right |,
$ and $y_{2i} = y_{2i-2}$. If a DLT was observed in the previous cohort of two patients and was attributable to drug $D_1$, then $x_{2i}$ is further restricted to be no more than $x_{2i-2}$.

\end{itemize}

\end{itemize}

\item Repeat step 2 until the maximum sample size is reached subject to the following stopping rule.

\item We would stop the trial if, $\mbox{Prob}( \mbox{Prob}(\mbox{DLT} | x=X_{\min}, y=Y_{\min}) \geq \theta + \xi_1 | \mbox{data}) > \xi_2$, i.e. if the posterior risk of toxicity at the lowest combination significantly is high. $\xi_1$ and $\xi_2$ are design parameters tuned to obtain the best operating characteristics.

\end{enumerate}

In step 2 of the algorithm, any dose escalation is
further restricted to be no more than a pre-specified fraction of the dose range of the corresponding agent. At the end of the trial, we obtain the MTD curve estimate $\hat{C} = C(\hat{\alpha},\hat{\beta}, \hat{\gamma})$, where $\hat{\alpha}$, $\hat{\beta}$ and $\hat{\gamma}$ are the posterior medians of the parameters $\alpha, \beta$ and $\gamma$, given the data.


\section{Simulation Studies}

\subsection{Simulation set up and Scenarios}
\label{simulationsetupscenarios}

In all simulated trials, the link functions $F_{\alpha}(x) = x^\alpha$ and $F_{\beta}(y) = y^\beta$ are used. To evaluate the performance of our proposal, the DLT outcomes are generated from the true model showed in (\ref{probdlt}). We used this model in 3 different scenarios to study the behavior of our design when the prior distribution of the model parameters is both well and poorly calibrated. Let $\alpha_{\scriptsize{\mbox{true}}}$, $\beta_{\scriptsize{\mbox{true}}}$ and $\gamma_{\scriptsize{\mbox{true}}}$ represent the true parameter values we use in (\ref{probdlt}) to generate DLT outcomes. In each scenario we select different values for $\alpha_{\scriptsize{\mbox{true}}}$, $\beta_{\scriptsize{\mbox{true}}}$, but the prior distribution for $\alpha$ and $\beta$, $P(\alpha)$ and $P(\beta)$, as well as $\gamma_{\scriptsize{\mbox{true}}}$, do not vary. In scenario 1, we choose values for $\alpha_{\scriptsize{\mbox{true}}}$ and $\beta_{\scriptsize{\mbox{true}}}$ such that $ \alpha_{\scriptsize{\mbox{true}}} < E[P(\alpha)]$ and $ \beta_{\scriptsize{\mbox{true}}} < E[P(\beta)]$. In scenario 2, we choose values for $\alpha_{\scriptsize{\mbox{true}}}$ and $\beta_{\scriptsize{\mbox{true}}}$ such that $ \alpha_{\scriptsize{\mbox{true}}} = E[P(\alpha)]$ and $\beta_{\scriptsize{\mbox{true}}} = E[P(\beta)]$. Last, in scenario 3, we choose values for $\alpha_{\scriptsize{\mbox{true}}}$ and $\beta_{\scriptsize{\mbox{true}}}$ such that $\alpha_{\scriptsize{\mbox{true}}} > E[P(\alpha)]$ and $ \beta_{\scriptsize{\mbox{true}}} > E[P(\beta)]$. Figure 2 shows the MTD curves with the true parameter values described here and their contours at $\theta \pm 0.05$ and $\theta \pm 0.1$.
We evaluate the effect of toxicity attribution in these 3 scenarios using 4 different values for $\eta$: 0, 0.1, 0.25 and 0.4. These values are reasonable because higher values of $\eta$ in practice are very rare. Data is randomly generated using the following procedure:

\begin{itemize}

\item For a given dose combination $(x,y)$, a binary indicator of DLT $T$ is generated from a Bernoulli distribution with probability of success computed using equation \eqref{probdlt}.

\item If $\{ T=1 \}$, we generate the attribution outcome $A$ using a Bernoulli distribution with probability of success $\eta$.

\item If $\{T=1, A=1\}$, we attribute the DLT to drug $D_1$, $D_2$, or to both drugs with equal probabilities.

\end{itemize}

We assume that the model parameters $\alpha, \beta, \gamma$ and $\eta$ are independent {\it a priori}. We assign vague prior distributions to $\alpha$, $\beta$ and $\gamma$ following \cite{yin2009latent}, where  $\alpha \sim \mbox{Uniform}(0.2,2)$, $\beta \sim \mbox{Uniform}(0.2,2)$ and $\gamma \sim \mbox{Gamma}(0.1,0.1)$. These prior distributions correspond to the ones used by \cite{yin2009latent} for the main analysis. The prior distribution for the fraction of attributable toxicities $\eta$ is set to be Uniform$(0,1)$. With these prior distributions, the true parameter values for each scenario are as follows. In scenario 1, $\alpha = \beta = 0.9$ and $\gamma = 1$. In scenario 2, $\alpha = \beta = 1.1$ and $\gamma = 1$. Last, in scenario 3, $\alpha = \beta = 1.3$ and $\gamma = 1$. For each scenario, $m=1000$ trials will be simulated. The target risk of toxicity is fixed at $\theta = 0.3$, the sample size is $n=40$, and the values for $\xi_1$ and $\xi_2$ will be 0.05 and 0.8 respectively. All simulation are done using the software R version 3.3.1.

\begin{figure}[H]
\caption{Contour plots for the working model in scenarios 1, 2 and 3. The black dashed curve represents the true MTD curve and the gray dashed lines represent the contours at $\theta \pm 0.05$ and $\theta \pm 0.10$.}
\centering
\includegraphics[scale=0.3]{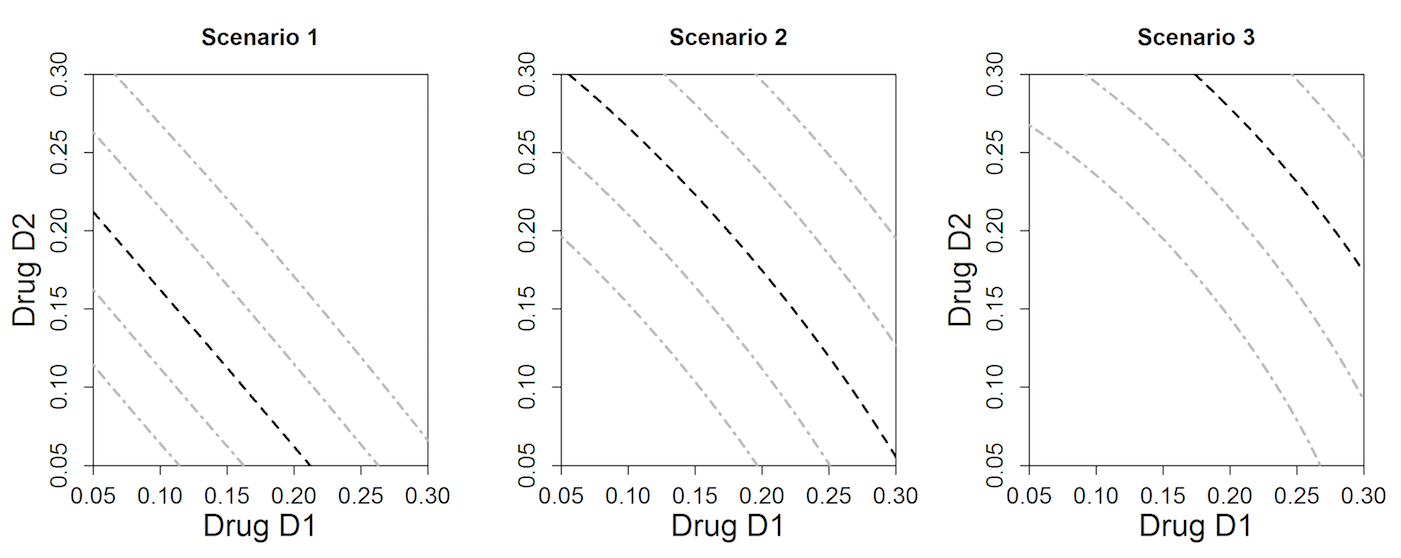}
\label{figure2}
\end{figure}

\subsection{Design Operating Characteristics}

We evaluate the performance of the design by assessing its safety and its efficiency in estimating the MTD curve.

For trial safety, we employ the average percent of DLTs, the percent of simulated trials with DLT rate greater than $\theta \pm 0.05$ and $\theta \pm 0.10$. 

For efficiency, we employ the pointwise average relative minimum distance from the true MTD curve to the estimated MTD curve. This measure of efficiency is well described in \cite{tighiouart2014dose,tighiouart2017bayesian} and can interpreted as a pointwise average bias in estimating the true MTD curve. We also consider the pointwise percent of trials for which the minimum distance of the point $(x,y)$ on the true MTD curve to the estimated MTD curve is no more than $(100 \times p) \%$ of the true MTD curve. This measurement will give us an estimate of the percent of trials with MTD recommendation within (100 × p)\% of the true MTD. This measure of efficiency can be interpreted as the pointwise percent of correct MTD recommendation. In this paper we select $p = 0.1, 0.2$. For a detailed explanation of these measures of efficiency, see \cite{tighiouart2014dose,tighiouart2017bayesian}.

\subsection{Results}

In general, increasing the value of $\eta$ until 0.4 generates estimated MTD curves closer to the true MTD curve. Figure 3 shows the estimated MTD curves for each scenario as a function of $\eta$. In terms of safety, overall we observe that increasing the fraction of toxicity attributions $\eta$ reduces the average percent of toxicities and percent of trials with toxicity rates greater than $\theta + 0.05$ and $\theta + 0.10$. Table \ref{safetycontinuous}, shows the average percent of toxicities as well as the percent of trials with toxicity rates greater than $\theta + 0.05$ and $\theta + 0.1$ for scenarios 1-3.

Figure 4 shows the pointwise average bias of the 3 proposed scenarios for each value of $\eta$. Overall, increasing the value of $\eta$ until 0.4 reduces the pointwise average bias. In any case, the pointwise average bias is around 10\% of the dose range of either drug and practically negligible for $\eta =0.25, 0.4$. For instance, under scenario 3, the maximum absolute value of the pointwise average bias when $\eta = 0.40$ is about 0.01, which corresponds to 0.3\% of the dose range, which is practically negligible.

Figure 5 shows the pointwise percent of MTD recommendation of the 3 proposed scenarios for each value of $\eta$. In general, increasing the value of $\eta$ increases the pointwise percent of MTD recommendation, reaching up to 80\% of correct recommendation when $p=0.2$, and up to 70\% of correct recommendation when $p=0.1$. Based on these simulation results, we conclude that in continuous dose setting the approach of partial toxicity attribution generates safe trial designs and efficient estimation of the MTD.

\begin{table}[H]
\caption{Operating characteristics summarizing trial safety in $m=1000$ simulated trials.}
\centering
\begin{tabular}{ccccc}
\toprule
 & & \makecell{Average \\ $\%$ of toxicities} & \makecell{$\%$ of trials with \\ toxicity rate $> \theta + 0.05$} & \makecell{$\%$ of trials with \\ toxicity rate $> \theta + 0.10$} \\ \midrule
\multirow{5}{*}{Scenario 1} & $\eta = 0.00$ & 33.62 & 25.90 & 4.10 \\ 
& $\eta = 0.10$ & 32.67  & 22.60 & 4.80 \\ 
& $\eta = 0.25$ & 31.55 & 17.60 & 2.70 \\ 
& \multicolumn{1}{l}{$\eta = 0.40$} & 30.70 & 13.30 & 2.00  \\ \midrule 
\multirow{5}{*}{Scenario 2} & $\eta = 0.00$ & 30.64  & 9.40 & 0.90 \\
& $\eta = 0.10$ & 29.69 & 7.30 & 0.40  \\ 
& $\eta = 0.25$ & 28.76 & 5.00 & 0.20 \\  
 & \multicolumn{1}{l}{$\eta = 0.40$} & 28.04 & 4.10 & 0.30  \\ \midrule 
\multirow{5}{*}{Scenario 3} & $\eta = 0.00$ & 27.47 & 2.00 & 0.00\\  
& $\eta = 0.10$ & 26.80 & 1.80 & 0.00 \\  
& $\eta = 0.25$ & 25.99 & 1.30 & 0.00 \\ 
 & \multicolumn{1}{l}{$\eta = 0.40$} & 25.37 & 0.70 & 0.00 \\ \bottomrule 
\end{tabular}
\label{safetycontinuous}
\end{table}

\begin{figure}[H]
\caption{Estimated MTD curves for $m=1000$ simulated trials. The black dashed curve represents the true MTD curve, the gray dashed lines represent the contours at $\theta \pm 0.05$ and $\theta \pm 0.10$, and the solid curves represent the estimated MTD curves at each value of $\eta$.}
\centering
\includegraphics[scale=0.3]{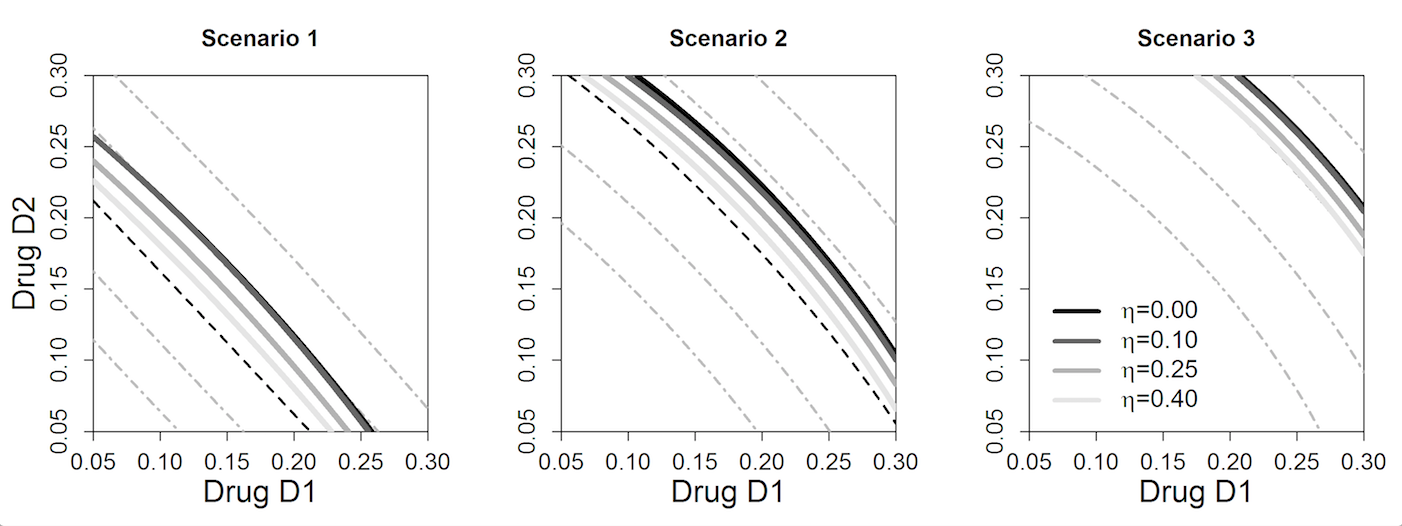}
\label{figure3}
\end{figure}

\begin{figure}[H]
\caption{Pointwise average bias in estimating the true MTD in $m=1000$ simulated trials.}
\centering
\includegraphics[scale=0.3]{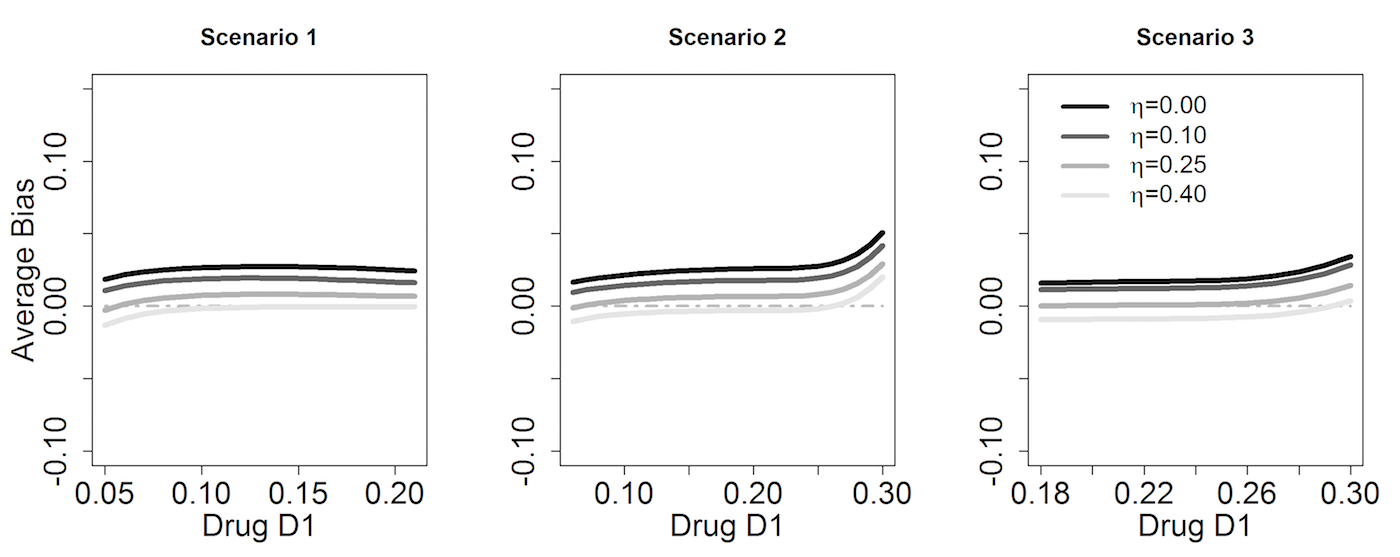}
\label{figure4}
\end{figure}

\begin{figure}[H]
\caption{Pointwise percent of MTD recommendation for $m=1000$ simulated trials. Solid lines represent the pointwise percent of MTD recommendation when $p=0.2$ and dashed lines represent the pointwise percent of MTD recommendation when $p=0.1$.}
\centering
\includegraphics[scale=0.3]{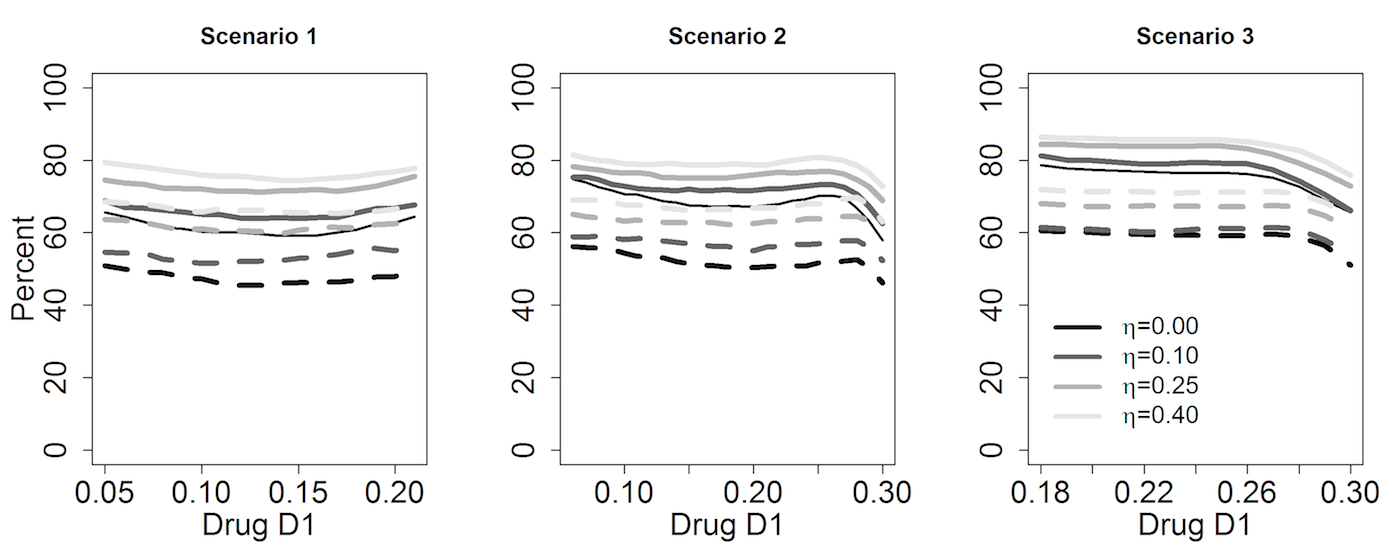}
\label{figure5}
\end{figure}


\section{Discrete Dose Combinations}

\subsection{Approach}

Dose escalation follows the same procedure described in section \ref{dosefindingalgorithmsection}. The only difference is that, in step 2, the continuous doses recommended are rounded to the nearest discrete dose level. For a detailed explanation of this procedure see \cite{tighiouart2017bayesian}.

\subsection{Illustration}
\label{illustrationsection}

We study the performance of our proposal in a discrete dose level setting where the probability of toxicity of each dose level is generated from the working model. We employ 6 scenarios with 4 dose levels respectively in each drug for scenarios 1 - 3, and 4 and 6 dose levels respectively in each drug for scenarios 4 - 6. The target probability of toxicity is always $\theta=0.3$ and, for each scenario, we simulate $m = 1000$ trials using the same vague priors for $\alpha$, $\beta$ and $\gamma$ specified in section \ref{simulationsetupscenarios}. The maximum sample size in all scenarios is again $n=40$. The performance of the method is evaluated using the percent of MTD selection statistic proposed by \cite{tighiouart2017bayesian}.

In Table \ref{toxicityscenarios} we present the 6 mentioned scenarios we use to illustrate the implementation of our design with discrete dose levels. Moreover, in Figure 6 we show the dose-toxicity surface of these 6 scenarios, where we observe that all of them have a flat (near-constant) surface.

\begin{figure}[H]
\caption{Probability of DLT surfaces of the 6 scenarios from Table 3.}
\centering
\includegraphics[scale=0.35]{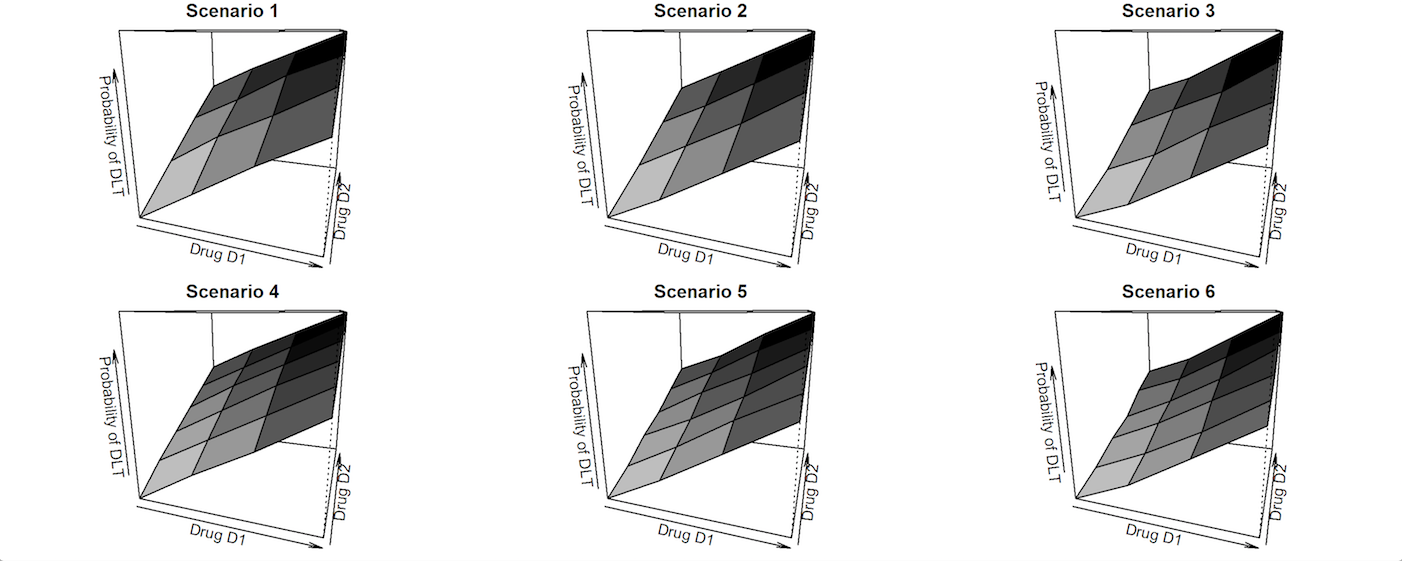}
\label{figure6}
\end{figure}

In Table \ref{percentrecommendedmtdset2} we show the percent of times that at least 25\%, 50\%, 75\% or 100\% of recommended MTDs belong to the true MTD set. Using vague prior distributions, the scenario where toxicity attribution has the strongest effect is scenario 2. In scenarios 1,4 and 5, we observe a slight effect but it does not make a big difference.

\begin{table}[H]
\caption{Dose limiting toxicity scenarios with $\theta=0.3$ generated with the working model. In bold the dose combination levels that would compose the true MTD set.}
\centering
\begin{tabular}{cccccccccccc}
\toprule
Dose level & 1      & 2      & 3     & 4     &  & 1    & 2    & 3    & 4    & 5    & 6    \\ \midrule
           & \multicolumn{4}{c}{Scenario 1} &  & \multicolumn{6}{c}{Scenario 4}         \\ 
4          & {\bf 0.39}   & 0.46   & 0.52  & 0.58  &  & {\bf 0.39} & 0.43 & 0.47 & 0.51 & 0.55 & 0.58 \\ 
3          & {\bf 0.31}   & {\bf 0.38}   & 0.46  & 0.52  &  & {\bf 0.30} & {\bf 0.35} & 0.40 & 0.44 & 0.48 & 0.52 \\ 
2          & {\bf 0.22}   & {\bf 0.31}   & {\bf 0.38}  & 0.46  &  & {\bf 0.22} & {\bf 0.27} & {\bf 0.32} & {\bf 0.37} & 0.41 & 0.46 \\ 
1          & 0.13   & {\bf 0.22}   & {\bf 0.31}  & {\bf 0.39}  &  & 0.13 & 0.19 & {\bf 0.24} & {\bf 0.29} & {\bf 0.34} & {\bf 0.39} \\  \midrule
           & \multicolumn{4}{c}{Scenario 2} &  & \multicolumn{6}{c}{Scenario 5}         \\ 
4          & {\bf 0.30}   & {\bf 0.36}   & 0.42  & 0.48  &  & {\bf 0.30} & {\bf 0.33} & {\bf 0.37} & 0.40 & 0.44 & 0.48 \\ 
3          & {\bf 0.22}   & {\bf 0.28}   & {\bf 0.35}  & 0.42  &  & {\bf 0.22} & {\bf 0.26} & {\bf 0.29} & {\bf 0.33} & {\bf 0.38} & 0.42 \\ 
2          & 0.14   & {\bf 0.21}   & {\bf 0.28}  & {\bf 0.36}  &  & 0.14 & 0.18 & {\bf 0.22} & {\bf 0.27} & {\bf 0.31} & {\bf 0.35} \\ 
1          & 0.07   & 0.14   & {\bf 0.22}  & {\bf 0.30}  &  & 0.07 & 0.11 & 0.16 & 0.20 & {\bf 0.25} & {\bf 0.30} \\ \midrule
           & \multicolumn{4}{c}{Scenario 3} &  & \multicolumn{6}{c}{Scenario 6}         \\ 
4          & {\bf 0.23}   & {\bf 0.27}   & {\bf 0.33}  & {\bf 0.39}  &  & {\bf 0.23} & {\bf 0.25} & {\bf 0.28} & {\bf 0.32} & {\bf 0.35} & {\bf 0.39} \\ 
3          & 0.16   & {\bf 0.21}   & {\bf 0.26}  & {\bf 0.33} &  & 0.16 & 0.18 & {\bf 0.22} & {\bf 0.25} & {\bf 0.29} & {\bf 0.33} \\ 
2          & 0.09   & 0.14   & {\bf 0.21}  & {\bf 0.27}  &  & 0.09 & 0.12 & 0.16 & 0.19 & {\bf 0.23} & {\bf 0.27} \\ 
1          & 0.04   & 0.09   & 0.16  & {\bf 0.23}  &  & 0.04 & 0.07 & 0.11 & 0.14 & 0.19 & {\bf 0.23} \\ \bottomrule
\end{tabular}
\label{toxicityscenarios}
\end{table}

\begin{table}[H]
\centering
\caption{Percent of times that at least 25\%, 50\%, 75\% or 100\% of recommended MTDs belong to the true MTD set in $m = 1000$ simulated trials.}
\resizebox{\columnwidth}{!}{%
\begin{tabular}{ccccccccccc}\toprule
&  \multicolumn{10}{c}{\makecell{$\%$ of correct MTD recommendation for $\theta \pm 0.10$ }} \\ \midrule
 & & $\geq 25\%$ & $\geq 50\%$ & $\geq 75\%$ & 100\%  &                           &$\geq 25\%$ & $\geq 50\%$ & $\geq 75\%$ &  100\%  \\ \midrule
$\eta = 0.00$ & \multirow{5}{*}{Scenario 1} & 91.40  & 87.30 & 83.70 & 83.70 &  \multirow{5}{*}{Scenario 4} &90.00  & 85.50 & {\bf 73.70} & 67.70  \\
$\eta = 0.10$ & & {\bf 92.50} & {\bf 87.80} & {\bf 83.90} & {\bf 83.90} &                           & {\bf 91.30} & {\bf 86.00} & 71.10 & 65.80   \\
$\eta = 0.25$ & & 90.90 & 87.70 & 83.80 & 83.80 &                           & 89.40 & 85.90 & 71.50 & {\bf 68.20}   \\
$\eta = 0.40$ & & 90.90 & 87.70 & 83.80 & 83.80 &                           & 89.80 & 85.70 & 70.50 & 67.30   \\ \midrule
$\eta = 0.00$ & \multirow{5}{*}{Scenario 2} & 78.10 & 78.10 & 73.60 & 73.60 & \multirow{5}{*}{Scenario 5} & 82.90 & 81.50 & {\bf 72.70} & {\bf 72.70} \\
$\eta = 0.10$ & & 79.80 & 79.80 & 73.90 & 73.90 &                           & 82.80 & 81.40 & 71.70 & 71.70   \\
$\eta = 0.25$ & & 83.00 & 83.00 & 75.50 & 75.50 &                           & 85.00 & 81.80 & 71.00 & 71.00   \\
$\eta = 0.40$ & & {\bf 83.50} & {\bf 83.50} & {\bf 76.20} & {\bf 76.20} &                           & {\bf 85.70} & {\bf 82.50} & 70.60 & 70.60   \\ \midrule
$\eta = 0.00$ & \multirow{5}{*}{Scenario 3} & 99.10 & {\bf 99.00} & {\bf 97.00} & {\bf 97.00} & \multirow{5}{*}{Scenario 6} & {\bf 98.80} & {\bf 96.50} & {\bf 93.10} & {\bf 92.70} \\
$\eta = 0.10$ & & {\bf 99.30} & 98.60 & 95.10 & 95.10 &                           & 98.60 & 95.70 & 89.90 & 89.60   \\
$\eta = 0.25$ & & 97.10 & 96.40 & 91.90 & 91.90 &                           & 96.20 & 92.10 & 87.00 & 86.20   \\
$\eta = 0.40$ & & 95.90 & 95.10 & 89.50 & 89.50 &                           & 94.00 & 89.50 & 81.90 & 81.00   \\ \bottomrule
\end{tabular}
}
\label{percentrecommendedmtdset2}
\end{table}


\section{Model Misspecification}

In the previous sections, all the simulated scenarios are generated with the model showed in (\ref{probdlt}). However, in practice we do not know the underlying model that generates the data and therefore we need to assess the performance of our design under model misspecification. We employ the same toxicity scenarios used by \cite{yin2009latent}, which are shown in Table \ref{toxicityscenarios2}. Moreover, In Figure 7 we show the dose-toxicity surface of these scenarios. Scenario 1 presents a very constant surface gradient. The rest of the scenarios present surface gradients that vary as we increase the dose combination levels. However, scenarios 3, 4 and 6 vary more abruptly than scenarios 2 and 5. Scenario 6 is a particular case because the lowest dose combination level has a probability of DLT that is already higher than the target risk of toxicity $\theta + 0.1$. Therefore, for this scenario, instead of presenting the  percent of correct recommendation we present the percent of times the trial is stopped due to safety using the stopping rule in Section 2.2 with $\xi_1 = 0.05$ and $\xi_2 = 0.8$. 
For each scenario, we simulate $m = 1000$ trials with a target risk of toxicity of $\theta=0.30$, a sample size of $n=40$ and we use the same prior distributions for $\alpha$, $\beta$ and $\gamma$ as in section \ref{simulationsetupscenarios}.

\begin{figure}[H]
\caption{Probability of DLT surfaces of the 6 scenarios from Table 5.}
\centering
\includegraphics[scale=0.35]{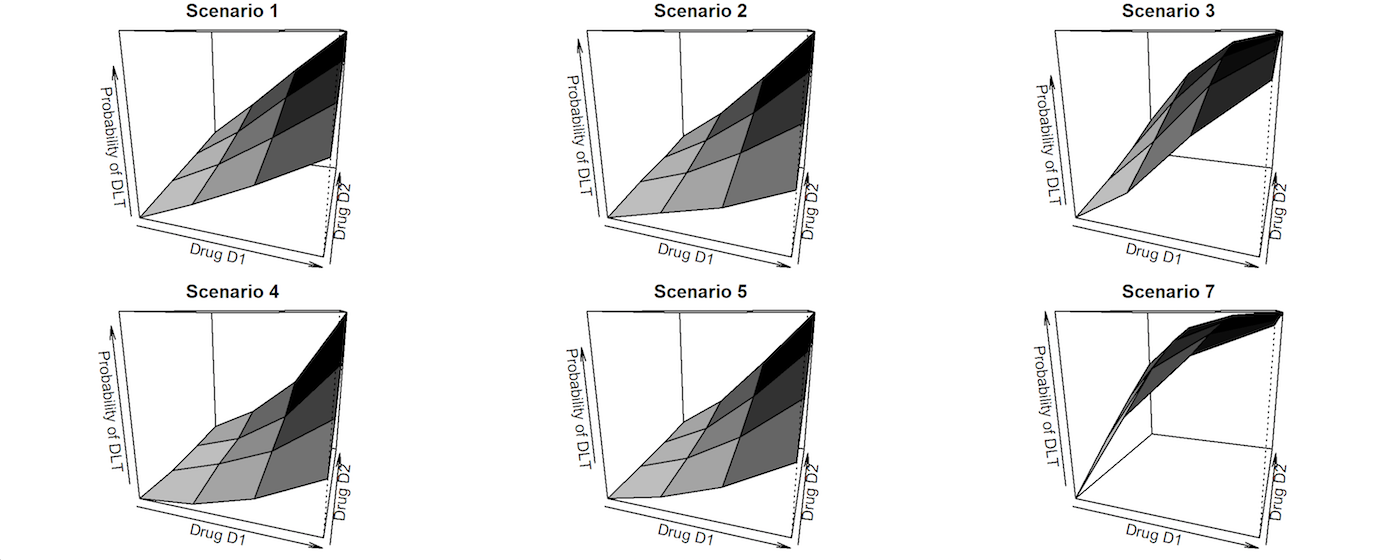}
\label{figure7}
\end{figure}

\begin{table}[t]
\caption{Toxicity scenarios for the model misspecification evaluation. In bold the doses considered as true MTD set.}
\centering
\begin{tabular}{cccccccccc}
\toprule
Dose level & 1      & 2      & 3     & 4     &  & 1    & 2    & 3    & 4   \\ \midrule
           & \multicolumn{4}{c}{Scenario 1} &  & \multicolumn{4}{c}{Scenario 4}         \\ 
4          &  {\bf 0.28}  &  0.41  & 0.55  & 0.68  &  & 0.04 & 0.09 & 0.17 & {\bf 0.32}   \\
3          & {\bf 0.25}  &  {\bf 0.35}  & 0.48  & 0.60   &  & 0.03 & 0.06 & 0.12 & {\bf 0.23}   \\
2          &  {\bf 0.22}  &  {\bf 0.30}  & {\bf 0.40}  & 0.51  &  & 0.02 & 0.05 & 0.09 & 0.16   \\
1          &  0.19  &  {\bf 0.26}  & {\bf 0.34}  & 0.43  &  & 0.02 & 0.03 & 0.06 & 0.11   \\ \midrule
           & \multicolumn{4}{c}{Scenario 2} &  & \multicolumn{4}{c}{Scenario 5}         \\ 
4          & 0.17   &  {\bf 0.29}  & 0.45  & 0.62  &  & 0.12 & {\bf 0.26} & 0.48 & 0.71  \\
3          & 0.14   & {\bf 0.23}   & {\bf 0.35}  & 0.50  &  & 0.09 & 0.19 & {\bf 0.36} & 0.57  \\
2          & 0.12   & 0.18   & {\bf 0.27}  & {\bf 0.38}  &  & 0.07 & 0.14 & {\bf 0.26} & 0.43  \\
1          & 0.09   & 0.14   & 0.19  & {\bf 0.27}  &  & 0.05 & 0.10 & 0.18 & {\bf 0.30}  \\ \midrule
           & \multicolumn{4}{c}{Scenario 3} &  & \multicolumn{4}{c}{Scenario 6}         \\ 
4          &  {\bf 0.37}  &  0.72  & 0.92  & 0.98  &  & 0.78 & 0.94 & 0.99 & 1.00 \\
3          &  {\bf 0.26}  &  0.59  & 0.85  & 0.96  &  & 0.68 & 0.90 & 0.97 & 0.99 \\
2          &  0.18  &  0.44  & 0.74  & 0.91  &  & 0.57 & 0.83 & 0.94 & 0.98 \\
1          &  0.12  &  {\bf 0.30}  & 0.59  & 0.82  &  & 0.45 & 0.73 & 0.90 & 0.97 \\ \bottomrule
\end{tabular}
\label{toxicityscenarios2}
\end{table}

In terms of safety, in general we observe that toxicity attributions reduce the average percent of toxicities and percent of trials with toxicity rates greater than $\theta + 0.05$ and $\theta + 0.10$. Table \ref{safetycontinuousmisspecification} shows the average percent of toxicities as well as the percent of trials with toxicity rates greater than $\theta + 0.05$ and $\theta + 0.1$.

In Table \ref{percentrecommendedmtdsetmissp2}, we show the percent of times that at least 25\%, 50\%, 75\% or 100\% of recommended MTDs belong to the true MTD set. In scenario 1 we observe a positive effect of the toxicity attributions, improving the percent of times at least 75\% and 100\% of recommended MTDs belong to the true MTD set in to 5\% when $\eta = 0.25$. In scenario 2 we observe a positive effect of the toxicity attributions improving the percent of times at least 25\% and 50\% of recommended MTDs belong to the true MTD up to 5\% and 4\% respectively. In scenarios 3, 4 and 5 we do not observe any positive effect when attributing toxicities. However, these scenarios are particularly difficult for our design given the rounding up procedure we follow with discrete dose combinations. In scenario 6 we observe a positive effect of the toxicity attributions, improving the percent of times the trial is stopped due to safety by almost 4\% when $\eta = 0.40$. 
Based on these simulation results under model misspecification, we conclude that the partial toxicity attribution method has good operating characteristics in recommending dose combinations of which, at least 50\% are the true MTDs; these percent of correct recommendations vary between 65\% to 98\% depending on the scenario. Moreover, there is a high probability of stopping the trial if there is evidence that the minimum dose combination in the trial has high probability of DLT.

However some of the scenarios showed in Table \ref{toxicityscenarios2} have a true set of MTDs that include a large number of dose combinations. For this reason, we implemented our design in 6 extra scenarios taken from \cite{yin2009latent,yin2009bayesian}. These scenarios are presented in Table S1 at the supplementary material, where the set of true MTDs contains a much more restricted number of dose combinations. Also, since the scenarios showed in Table \ref{toxicityscenarios2} where generated with a logistic model, we selected the scenarios to observe how robust is our proposal in scenarios generated with other models, such us the Clayton Copula, and scenarios that are arbitrarily generated. Moreover, since we are using the same set of true MTDs as \cite{yin2009latent,yin2009bayesian}, we use these methods to make a performance comparison in terms of percent of correct MTD selection.

In Tables S2 and S3, in the supplementary material, we present operating characteristics in term of safety and efficiency for each of the 6 proposed scenarios. In general, we observe that the design behaves in a similar way as with the scenarios presented along this manuscript. In terms of safety, toxicity attributions reduce the average percent of toxicities and the percent of trials with toxicity rates greater than $\theta + 0.05$ and $\theta + 0.10$. In terms of efficiency, we only observe a positive effect in scenarios with a relatively flat dose-toxicity surface. In terms of performance comparison, our proposed method is competitive with other standard designs for drug combinations such us \cite{yin2009latent,yin2009bayesian}, and achieves better percent of correct MTD recommendation in 4 out of the 6 used scenarios.

Another issue that is relevant to the methodology we present in this manuscript is the errors in the attribution of toxicities by the treating investigators. Our design does not include a parameter to control the uncertainty around the decision made by the investigator when attributing the the DLT, which could be an extension of this work. However, in the supplementary material, in order to assess the impact of these kind of errors, we present simulation from 3 scenarios taken also from \cite{yin2009latent,yin2009bayesian} where we introduce 10\% and 50\% of errors in the attribution of DLTs, and compare it to the case where we correctly attributes 100\% of the DLTs. In Tables S4 and S5, we present the simulated results in terms of safety and efficiency. Overall we do not observe any major difference when incorrectly attributing 10\% and 50\% of the DLTs with respect to correctly attributing 100\% of the DLTs.

\begin{table}[t]
\caption{Operating characteristics summarizing trial safety for model misspecification in $m=1000$ simulated trials.}
\centering
\begin{tabular}{ccccc}
\toprule
 & & \makecell{Average \\ $\%$ of toxicities} & \makecell{$\%$ of trials with \\ toxicity rate $> \theta + 0.05$} & \makecell{$\%$ of trials with \\ toxicity rate $> \theta + 0.10$} \\ \midrule
\multirow{5}{*}{Scenario 1} & $\eta = 0.00$ & 32.99 & 22.90 & 4.20 \\ 
& $\eta = 0.10$ & 32.19  & 18.50  & 2.90  \\ 
& $\eta = 0.25$ & 31.43 & 15.80 & 2.60  \\ 
& \multicolumn{1}{l}{$\eta = 0.40$} & 30.58 & 12.90 & 2.50  \\ \midrule 
\multirow{5}{*}{Scenario 2} & $\eta = 0.00$ & 29.85  & 6.60  & 0.20 \\ 
& $\eta = 0.10$ & 29.14  & 4.10  & 0.10 \\ 
& $\eta = 0.25$ & 28.20 & 3.10 & 0.30  \\ 
& \multicolumn{1}{l}{$\eta = 0.40$} & 27.90 & 2.30 & 0.00  \\ \midrule 
\multirow{5}{*}{Scenario 3} & $\eta = 0.00$ & 36.53 & 40.70  & 16.40  \\ 
& $\eta = 0.10$ & 35.13 & 33.90  & 12.50 \\ 
& $\eta = 0.25$ & 33.94  & 28.60 & 11.00  \\ 
& \multicolumn{1}{l}{$\eta = 0.40$} & 32.94 & 23.40 & 9.50  \\ \midrule 
\multirow{5}{*}{Scenario 4} & $\eta = 0.00$ & 22.43 & 0.00  & 0.00 \\ 
& $\eta = 0.10$ &  21.83  & 0.00 & 0.00 \\ 
& $\eta = 0.25$ & 21.39  & 0.00  & 0.00  \\ 
& \multicolumn{1}{l}{$\eta = 0.40$} & 20.87 & 0.00 & 0.00  \\ \midrule 
\multirow{5}{*}{Scenario 5} & $\eta = 0.00$ & 30.43  & 6.60  & 0.30 \\ 
& $\eta = 0.10$ & 29.48  & 3.30 & 0.10 \\ 
& $\eta = 0.25$ & 28.60  & 3.60 & 0.00 \\ 
& \multicolumn{1}{l}{$\eta = 0.40$} & 27.60 & 2.30 & 0.00  \\ \bottomrule 
\end{tabular}
\label{safetycontinuousmisspecification}
\end{table}

\begin{table}[t]
\centering
\caption{Percent of times that at least 25\%, 50\%, 75\% or 100\% of recommended MTDs belong to the true MTD set in $m = 1000$ simulated trials under model misspecification. In scenario 6 we show the percent of times the trial is stopped due to safety reasons.}
\resizebox{\columnwidth}{!}{%
\begin{tabular}{ccccccccccc}\toprule
&  \multicolumn{10}{c}{\makecell{$\%$ of correct MTD recommendation for $\theta \pm 0.10$ }} \\ \midrule
 & & $\geq 25\%$ & $\geq 50\%$ & $\geq 75\%$ & 100\%  &                           &$\geq 25\%$ & $\geq 50\%$ & $\geq 75\%$ &  100\%  \\ \midrule
$\eta = 0.00$ & \multirow{4}{*}{Scenario 1} & 82.90 & 75.60 & 55.40 & 55.40 &  \multirow{4}{*}{Scenario 4} & {\bf 98.80} & {\bf 98.80} & {\bf 87.80} & {\bf 87.80}  \\
$\eta = 0.10$ & &82.70 & 72.70 & 57.30 & 57.30 &                           & 97.20 & 97.20 & 85.70 & 85.70  \\
$\eta = 0.25$ & & {\bf 83.30} & {\bf 75.80} & {\bf 60.40} & {\bf 60.40} &                           & 95.70 & 95.70 & 82.00 & 82.00  \\
$\eta = 0.40$ & & 80.60 & 73.10 & 57.60 & 57.60 &                           & 95.20 & 95.20 & 76.20 & 76.20   \\ \midrule
$\eta = 0.00$ & \multirow{4}{*}{Scenario 2} &74.70 & 71.00 & {\bf 58.20} & {\bf 45.70} & \multirow{4}{*}{Scenario 5} &   {\bf 75.40} & {\bf 69.10} & {\bf 20.50} & {\bf 20.50} \\
$\eta = 0.10$ & & 77.00 & 73.50 & 53.60 & 44.60 &                           & 71.80 & 62.80 & 20.40 & 20.40 \\
$\eta = 0.25$ & & {\bf 79.60} & {\bf 75.00} & 50.00 & 41.20 &            & 70.70 & 59.40 & 18.90 & 18.90 \\
$\eta = 0.40$ & & 77.30 & 73.10 & 47.90 & 37.80 &                           & 71.20 & 60.50 & 16.70 & 16.70 \\ \midrule
$\eta = 0.00$ & \multirow{4}{*}{Scenario 3} & {\bf 76.90} & {\bf 65.30} & {\bf 23.30} & {\bf 23.30} & \multirow{4}{*}{Scenario 6} &\multicolumn{4}{c}{83.60}                              \\
$\eta = 0.10$ & & 72.50 & 61.80 & 21.90 & 21.90 &                             & \multicolumn{4}{c}{82.90}            \\
$\eta = 0.25$ & & 66.40 & 57.30 & 18.60 & 18.60 &                             & \multicolumn{4}{c}{84.80}                              \\
$\eta = 0.40$ & & 66.10 & 54.70 & 15.70 & 15.70  &                             & \multicolumn{4}{c}{{\bf 87.20}}   \\ \bottomrule
\end{tabular}
}
\label{percentrecommendedmtdsetmissp2}
\end{table}

\section{Conclusions}

In this paper we proposed a Bayesian adaptive design for cancer phase I clinical trials using drug combinations with continuous dose levels and attributable DLT in a fraction of patients. A copula-type model was used to describe the relationship between dose combinations and probability of DLT. The trial design proceeds by treating cohorts of two patients, each patient with a different dose combination estimated using univariate CRM for a better exploration of the space of doses. Treating cohorts of two patients will allow trial conduct to be completed in a reasonable amount of time. Although the two patients in a cohort are allocated to different dose combinations, a patient in the current cohort can be treated at a dose $(x, y)$ if and only if a patient in the previous cohort was treated at a dose on the same horizontal or vertical line within our dose range, that is was treated with either dose $x$ or dose $y$. The use of continuous dose levels is not uncommon in early phase trials, particularly when the drugs are given as infusions intravenously. For instance, a drug combination trial of cabazitaxel and cisplatin delivered intravenously was recently designed for advanced prostate cancer patients where the dose levels are continuous and the protocol was approved by the scientific review at Cedars-Sinai. For ethical reasons, we further imposed dose escalation restrictions for one of the drugs when a DLT is attributable to that drug.  

We studied the operating characteristics of the design under various scenarios for the true location of the MTD curve. In general, we observed that the trial is safe and as the proportion of attributed toxicities increases, the average proportion of toxicities decreases when we attribute toxicities. To assess the efficiency when estimating the MTD curve, we employed the pointwise average bias and average percent selection. In general the method is efficient although the results varied depending on the proportion of attributed toxicities. Note that the operating characteristics were evaluated under vague prior distributions of the model parameters and no toxicity profiles of single agent trials were used \emph{a priori}. We also showed how the method can be adapted to the setting of discrete dose combinations. 

We also performed a model misspecification evaluation in scenarios with different dose-toxicity surfaces. We only observed a positive effect in terms of percent of correct MTD recommendation in scenarios with flat surfaces. In scenarios with non-flat dose-toxicity surfaces we observed a decline in performance of percent selection consistent with the findings by \cite{riviere2014bayesian} when working with copula regression models. We also observed a positive effect in scenarios where the lowest dose combination has an excessively high probability of DLT. In this case, toxicity attributions improves the percent of times the trial was stopped due to safety. In all cases, safety of the trial is not compromised by accounting for a partial toxicity attribution. Clearly, there is a trade-off when increasing the fraction of DLT attribution to one or more drugs. The design is more conservative in future escalations, lowering the in-trial DLT percentages and reducing how quickly the MTD contour is reached, by favoring experimentation over recommendation.

Our design is practically useful when the two drugs do not have many overlapping toxicities, see e.g. \cite{miles2002combination} for some examples of drug combination trials with these characteristics. In cases where we expect a high percent of overlapping DLTs, designs that do not distinguish between drug attribution listed in the introduction may be more appropriate. Our method relies on clinical judgment regarding DLT attribution. In many phase I trials, such decisions are subject to error classifications and a possible extension is to introduce a parameter to account for errors in toxicity attribution as in \cite{iasonos2016phase} for single agent trials. We also plan to study the performance of this design using other link functions under different copula models, and extend this method to early phase cancer trials with late onset toxicity and by accounting for patient's baseline characteristic by extending the approaches in \cite{tighiouart2014escalation, tighiouart2012covariate} to the drug combination setting. 

\section{Acknowledgments}

This project has received funding from the European Union's Horizon 2020 research and innovation programme under the Marie Sklodowska-Curie grant agreement No 633567 (J.J. and M.G.), the National Institute of Health Grant Number 1R01CA188480-01A1 (M.T.), the National Center for Research Resources, Grant UL1RR033176, and is now at the National Center for Advancing Translational Sciences, Grant UL1TR000124 (M.T.), and 2 P01 CA098912 (M.T.).

\section{Conflict of Interests Statement}
The authors have declared no conflict of interest.

\newpage{}

\section*{Supplementary Material}

The supplementary material can be found at \\
 \texttt{https://onlinelibrary.wiley.com/doi/10.1002/bimj.201700166}.

\end{document}